\documentclass[journal,twoside,twocolumn]{IEEEtran}
\pdfoutput=1  %

\usepackage{amsmath,amssymb}
\usepackage{graphicx}
\usepackage{xcolor}
\usepackage{cite}
\usepackage{balance}
\usepackage{hyperref}
\usepackage[normalem]{ulem} %

\hypersetup{
  breaklinks=true,  %
  colorlinks=true,  %
  pdfusetitle=true,  %
  citecolor=blue,  %
  linkcolor=blue  %
}

\newcommand{\R}{{\mathbb{R}}}
\newcommand{\Z}{{\mathbb{Z}}}
\newcommand{\T}{{\mathrm{T}}}
 
\newcommand{\bA}{{\boldsymbol{A}}}
\newcommand{\bB}{{\boldsymbol{B}}}
\newcommand{\bg}{{\boldsymbol{g}}}
\newcommand{\bu}{{\boldsymbol{u}}}
\newcommand{\bx}{{\boldsymbol{x}}}
\newcommand{\bz}{{\boldsymbol{z}}}
\newcommand{\bzero}{{\boldsymbol{0}}}

\newcommand{\Ldual}{L^*}
\newcommand{\h}{\tfrac{1}{2}} %
\newcommand{\dx}{\,\mathrm{d}\bx}

\newcommand{\eqlab}[2]{\begin{align} \label{#1} #2 \end{align}}
\newcommand{\eq}[1]{\begin{align} #1 \end{align}}

\title{Glued lattices are better quantizers than $K_{12}$ }
\author{Erik Agrell, \IEEEmembership{Fellow, IEEE}, Daniel Pook-Kolb, and Bruce Allen, \IEEEmembership{Member, IEEE}
\thanks{Manuscript received 11 December 2023; accepted 29 April 2024 without revision.
The work of E.~Agrell was supported by a Collaborating Scientist Grant from the Max Planck Institute for Gravitational Physics, Germany, which is gratefully acknowledged.}
\thanks{E.~Agrell is with the Department of Electrical Engineering, Chalmers University of Technology, 41296 Gothenburg, Sweden (e-mail: agrell@chalmers.se).}
\thanks{D.~Pook-Kolb and B.~Allen are with the Max Planck Institute for Gravitational Physics, 30167 Hannover, Germany, and also with the Institut f\"ur Gravitationsphysik, Leibniz Universit\"at Hannover, 30167 Hannover, Germany (e-mail: daniel.pook.kolb@aei.mpg.de, bruce.allen@aei.mpg.de).}
}

\begin{document}

\maketitle

\begin{abstract}
  40 years ago, Conway and Sloane proposed using the highly symmetrical
  Coxeter--Todd lattice $K_{12}$ for quantization, and estimated its
  second moment. Since then, all published lists identify $K_{12}$ as
  the best 12-dimensional lattice quantizer. Surprisingly, $K_{12}$ is
  not optimal: we construct two new 12-dimensional lattices with lower
  normalized second moments. The new lattices are obtained by
  gluing together products of two 6-dimensional lattices.
\end{abstract}


\section{Introduction}

\IEEEPARstart{O}{ne} of the classical problems in lattice theory is to
find the best \emph{lattice quantizer,} i.e., the lattice with minimum
normalized second moment in a given dimension
\cite[Ch.~2]{conway99splag}. This problem has applications in data
compression \cite{gray98}, \cite[Chs.~1, 3--5]{zamir14book}, geometric
shaping of modulation formats \cite{forney89a},
\cite[Ch.~9]{zamir14book}, coding for noisy channels
\cite[p.~70]{conway99splag}, experimental design \cite{hamprecht03},
and data analysis \cite{allen21}.

The lattice quantizer problem was pioneered by Fejes T\'oth, who
showed that the hexagonal lattice is optimal in two dimensions
\cite{fejestoth59}. The corresponding optimum in three dimensions is
the body-centered cubic lattice, as proved by Barnes and Sloane in
1983 \cite{barnes83}. In higher dimensions $n$, the optimal lattices
are not known. Tables of the best known lattices were presented for
$n\le 5$ in \cite{gersho79}, $n\le 10$ in \cite{conway82voronoi,
  agrell98}, $n=1,\ldots,8,12,16,24$ in \cite{conway84}, \cite[pp.~12,
  61]{conway99splag}, \cite[p.~135]{zamir14book}, $9\le n \le 12$ in
\cite{dutour09}, $n=1,\ldots,10,12,16,24$ in \cite{torquato10}, $n\le
15$ in \cite{allen21}, $n\le 24$ in \cite{lyu22}, and $n\le 48$ in
\cite{agrell23}.

The above-mentioned tables devote more attention to some dimensions
than others. One such dimension is $n=12$, and the reason for the
interest in this dimension is the existence of the highly symmetrical
\emph{Coxeter--Todd lattice} $K_{12}$, which was discovered by Coxeter
and Todd in 1953 \cite{coxeter53}.
Its second moment was
computed numerically in 1984 \cite{conway84}
and exactly in 2009 \cite{dutour09}.
It is listed as the best
$12$-dimensional lattice quantizer in all tables we have seen. The
lattice was conjectured optimal for both quantization and packing in
\cite[p.~13]{conway99splag}. Its construction, symmetry group, and
other properties are described in \cite{conway83mpcps},
\cite[pp.~127--129]{conway99splag}.

In this paper, we prove that $K_{12}$, contrary to the popular belief, is
\emph{not} the optimal lattice quantizer in 12 dimensions. We do this
by designing two better lattices and computing their second moments
exactly. These new lattices are constructed using \emph{gluing
theory,} which was introduced by Conway \emph{et al.} for self-dual block codes
in \cite{conway79} and for integral lattices in
\cite{conway82jnt}. Using this theory, self-dual integral lattices
were constructed as the union of a finite number of translates of a
given base lattice, which is typically a \emph{product lattice}
\cite{conway82ejc}, \cite[Sec.~3 in Ch.~4]{conway99splag}. Gluing
or similar techniques have not, as far as we know, been applied
in the quest for good lattice quantizers, which are not
necessarily integer lattices. That is the scope of the present paper.

\section{Lattice Fundamentals} \label{s:fundamentals}

A \emph{lattice} $L$ is an infinite, countable set of real vectors
that forms a finitely generated group under addition.
For $n \le m$, it can be defined by
an $n\times m$ real \emph{generator matrix} $\bB$ with linearly
independent rows, such that
\eq{
L = \{\bu \bB \colon \bu \in \Z^n\}
,}
where $\bu$ are $n$-dimensional row vectors of integers.  The
rows of $\bB$ are called \emph{basis vectors.} The \emph{dimension} of
$L$ is $n$, and the lattice is embedded in the Euclidean space
$\R^m$. There exist infinitely many generator matrices for the same
lattice.

The $n \times n$ symmetric, positive definite \emph{Gram matrix} $\bA
\triangleq \bB \bB^\T$ gives the inner products of all basis vectors
with each other. There exist infinitely many Gram matrices for the same lattice,
but their determinants are equal, and this value is called the
\emph{lattice determinant} \cite[p.~4]{conway99splag}.

The lattice generated by $\bB^* \triangleq \bA^{-1} \bB$, which is the transpose
of the pseudoinverse of $\bB$, is the \emph{dual lattice} $\Ldual$.
The Gram matrix of this dual lattice
is $\bA^{-1}$. If $m=n$, then $\bA^{-1} = {(\bB^\T)}^{-1} \bB^{-1}$
and $\bB^* = (\bB^\T)^{-1}$.  If $L=\Ldual$, then the lattice is said
to be \emph{self-dual}.

The \emph{Voronoi region} of the lattice generated by $\bB$ is
\eq{
  \Omega \triangleq \Big\{\bz \bB \colon \bz \in \R^n, \min_{\bu \in \Z^n}
  \| \bz \bB - \bu \bB\|^2 = \| \bz \bB \|^2 \Big\}
.}
This is the set of
all vectors (in the space spanned by the lattice vectors) whose
closest lattice vector is the all-zero vector $\bzero$. The Voronoi
region of any lattice is a convex polytope, which is symmetric under
reflection through $\bzero$ and has volume $V=\sqrt{\det\bA}$.
The
facets of the Voronoi region lie in $(n-1)$-dimensional planes that are
equidistant from $\bzero$ and another lattice vector; these lattice
vectors are called \emph{relevant vectors.}
The vertices of the
Voronoi region are called \emph{holes} of the lattice, and the
vertices farthest away from $\bzero$ are called \emph{deep holes.}

For many purposes, it is desirable that the Voronoi region is as
spherical as possible according to some criteria. For lattice
quantizers, we seek to minimize the second moment (moment of inertia)
of the Voronoi region. Using the customary normalization, the relevant
figure of merit is the \emph{normalized second moment} (NSM) or
\emph{quantizer constant}
\eq{
  G = \frac{1}{nV^{1+2/n}}\int_\Omega \|\bx\|^2 \dx
 .}
 The normalization by $V^{1+2/n}$ makes $G$
invariant under rescaling, and normalization by $n$ ensures that a
product lattice (defined below) of identical lattices $L$ has the same
NSM as $L$.

The (Cartesian) \emph{product} of two lattices (or any other vector
sets) is
\eq{ L_1 \times L_2 \triangleq \{ [\bx_1 \; \bx_2] \colon
  \bx_1 \in L_1, \bx_2 \in L_2 \}
.}
The Voronoi region of a product
lattice is the product of the Voronoi regions of each of its
lower-dimensional component lattices. Hence, the set of holes of a
product lattice is the Cartesian product of the set of holes of each
component lattice. The relevant vectors of a product lattice are
$[\bx_1 \, \bzero]$ and $[\bzero \, \bx_2]$,
where $\bx_i$ are the relevant vectors of $L_i$.  Thus, the number of
facets of the product lattice is the sum of the number of facets of
the component lattices.
Product lattices are the currently best known lattice quantizers in
many dimensions \cite{gersho79, lyu22, agrell23},
but they are not optimal, because their NSM can always be decreased by
small perturbations of the generator matrix 
\cite[Th.~7]{agrell23}.  This
motivates a search for systematic ways to improve product lattices. The approach
we follow here is provided by gluing theory.

\section{Gluing Theory}

Gluing theory was developed in the context of the well-studied
\emph{integer lattices,} i.e., lattices whose Gram matrix contains
only integers. An interesting feature is that an integer lattice $L$
is always a sublattice of its dual $\Ldual$. This means that $\Ldual$
can be written as the union of $L$ and a finite number of its
translates
\eqlab{e:ldual}{
  \Ldual = \bigcup_{\bg \in \Ldual/L} (L+\bg).
}
Here $\Ldual/L$ denotes a (finite) set of coset
representatives of $L$ in $\Ldual$ \cite[p.~48]{conway99splag}.

These coset representatives are called \emph{glue vectors} of $L$. The
number of glue vectors is $\det\bA$, and by convention we use $\bzero$
as the coset representative for $L$ itself, so $\bzero$ is always a glue
vector.  Glue vectors of the root lattices $A_n$, $D_n$, and $E_n$ are
listed in \cite[Tab.~I]{conway82jnt}, \cite[Ch.~4]{conway99splag}.
We
say that a vector $\bx\in \Ldual$ is of \emph{type} $\bg$ if $\bx \in
L+\bg$.

If some glue vectors are omitted from the union in \eqref{e:ldual},
then a more general construction \cite{gannon91}
\eqlab{e:ltilde}{
  \tilde{L} \triangleq \bigcup_{\bg \in \Gamma} (L+\bg)
}
is obtained,
where $\{\bzero\} \subseteq \Gamma \subseteq \Ldual/L$.
We call this construction ``gluing.'' The term was introduced in
\cite{conway79, conway82jnt, conway82ejc}
for the special case when $L$ is a product code or product lattice;
as in \cite{gannon91}, we use it more generally for any lattice constructed by \eqref{e:ltilde}.
In one
extreme, \eqref{e:ltilde} yields $L$ and in the other extreme
$\Ldual$. Intermediate choices of $\Gamma$ can yield interesting
families of lattices or nonlattice packings, such as the Coxeter
lattices $A_n^r$ \cite{coxeter51}, \cite[Sec.~5.1]{dutour09} and
$D_n^+$ \cite[p.~119]{conway99splag}.
The so-called \emph{Construction A} can be seen as a special case
of \eqref{e:ltilde} with $L=\sqrt{2}\Z^n$  and $\Gamma$ being a rescaled
binary block code \cite[pp.~137--141, 182--185]{conway99splag}.

$\tilde{L}$ is a lattice if and only if $\Gamma$ is a group under
addition modulo $L$. If so, $\Gamma$ is called the \emph{glue group}
of $\tilde{L}$, and a generator matrix for $\tilde{L}$ can be obtained
as follows.
Starting with a generator matrix for $L$, we append the elements of $\Gamma$ 
as $|\Gamma|$ additional rows. Then we carry out linear row transformations 
with integer coefficients to make $|\Gamma|$ rows all-zero, and remove those 
all-zero rows.

The Voronoi region of $\tilde{L}$ is contained in the Voronoi region
of $L$. Specifically, if some elements of $\Gamma$ are located in holes of $L$,
then the corresponding vertices of the Voronoi region are ``cut away.''
This intuitively explains why gluing can potentially make Voronoi regions
more spherical and create better lattice quantizers.

We exploit this construction technique, selecting $L$ as the product
$L = L_1 \times \cdots \times L_k$ of known integer lattices
$L_i$. Then $L$ is also an integer lattice and its glue vectors are
Cartesian products of the glue vectors of $L_1,\ldots,L_k$.  When
written as row vectors, this concatenates them, so the glue vectors of
a product lattice $L$ are called \emph{glue words.}

When $L$ is a product lattice,
the set of glue words in \eqref{e:ldual} is the Cartesian product of
the sets of glue vectors of the component lattices, $\Ldual/L =
(L_1^*/L_1) \times \cdots \times (L_k^*/L_k)$. There are
$(\det\bA_1)\cdots(\det\bA_k)$ such glue words, where $\bA_i$ is a
Gram matrix of $L_i$, and hence many options when constructing new
lattices via \eqref{e:ltilde}.

Conway and Sloane studied a large number of glued lattices $\tilde{L}$
generated from various product lattices $L$ by \eqref{e:ltilde}. Their
goal was to construct integer lattices with determinant one, which are
self-dual. They successfully enumerated the lattice components and
glue words of all such lattices in dimensions up to $24$
\cite{conway82jnt,conway82ejc}. In the next section, we use the same
technique to find better lattice quantizers.
 
More generally, the construction \eqref{e:ltilde} is also valid when
$\Gamma$ is not a subset of $L^*/L$. For example, the
$9$-dimensional lattice quantizer with the smallest known NSM
is $L \cup (L+\bg)$ (corresponding to $\Gamma=\{\bzero,\bg\}$),
where $L=D_8 \times 2a\Z$, $\bg$ is a deep hole of $L$, and
$a$ is an algebraic scalar constant \cite{agrell98,allen21}. In
this case, $L$ is not an integer lattice and $L^*/L$ does not exist.
While interesting, such generalizations of \eqref{e:ltilde} are not
considered further in this paper.

\section{New Lattices}

\begin{table*}
\caption{Lattices $\tilde{L}$ generated by gluing $L = E_6\times E_6$ according to \eqref{e:ltilde}.}
\label{t:e6e6}
\begin{center}
\renewcommand\arraystretch{1.3}  %
\begin{tabular}{llll}
\hline
Glue words $\Gamma$ & Estimated NSM of $\tilde{L}$ & Exact NSM & Comment \\
\hline\hline
$\{\bg_{00}\}$                  & $0.074336 \pm 0.000010$ & $G_{E_6} \approx 0.074347$         & $\tilde{L} = E_6\times E_6$ \\
$\{\bg_{00},\bg_{01},\bg_{02}\}$ & $0.075557 \pm 0.000010$ & $7711/102\,060 \approx 0.075554$  & $\tilde{L} = E_6\times E_6^*$ \\
$\{\bg_{00},\bg_{10},\bg_{20}\}$ & $0.075556 \pm 0.000010$ & Same as the previous              & $\tilde{L} = E_6^*\times E_6$, equivalent to the previous \\
$\{\bg_{00},\bg_{11},\bg_{22}\}$ & $0.070060 \pm 0.000007$ & Given by \eqref{e:nsme6e6}             & $\tilde{L}$ is a better quantizer than $K_{12}$! \\
$\{\bg_{00},\bg_{12},\bg_{21}\}$ & $0.070063 \pm 0.000007$ & Same as the previous              & Equivalent to the previous \\
$\{\bg_{00},\bg_{01},\bg_{02},\bg_{10},\bg_{11},\bg_{12},\bg_{20},\bg_{21},\bg_{22}$\}            & $0.074237 \pm 0.000009$ & $G_{E_6^*} \approx 0.074244$ & $\tilde{L} = E_6^*\times E_6^*$ \\
\hline
\end{tabular}
\end{center}
\end{table*}

\subsection[]{Glued $E_6\times E_6$} \label{s:e6e6}

The integer lattice $E_6$ is often defined as a sublattice of the
\emph{Gosset lattice} $E_8$. This approach gives rise to the
rectangular generator matrix in \cite[p.~126]{conway99splag}, where
$m=8$. We find it more convenient to work with the square generator
matrix
\eq{
\left[\begin{array}{cccccc}
2 & 0 & 0 & 0 & 0 & 0 \\
1 & 1 & 0 & 0 & 0 & 0 \\
1 & 0 & 1 & 0 & 0 & 0 \\
1 & 0 & 0 & 1 & 0 & 0 \\
1 & 0 & 0 & 0 & 1 & 0 \\
\h & \h & \h & \h & \h & \tfrac{\sqrt{3}}{2}
\end{array}\right]
,}
which is obtained by applying linear row operations to
\cite[Eq.~(38)]{agrell98} and negating the sign of one column.

Since $\det\bA = (\det\bB)^2 = 3$, the lattice has three glue vectors
$E_6^*/E_6$. These are given in the traditional $8$-dimensional
representation in \cite[p.~126]{conway99splag}. The corresponding
glue vectors in our $6$-dimensional representation are
\eq{
\begin{array}{c@{\,}c@{}cc@{}cc@{}cc@{}cc@{}cc@{}c@{\,}c}
\bg_0=[ && 0 && 0 && 0 && 0 && 0 && 0 & ], \\
\bg_1=\big[ &-&\h &-&\h &-&\h &-&\h &-&\h &&\tfrac{\sqrt{3}}{6} & \big], \\
\bg_2=\big[ &&\h &&\h &&\h &&\h &&\h &-&\tfrac{\sqrt{3}}{6} & \big] .
\end{array}
}
The (Abelian) glue group is the cyclic group on 3 elements $Z_3$.
It has $\bg_0$ as the identity, and multiplication table $\bg_1^2 =
\bg_2$, $\bg_2^2 = \bg_1$, and $\bg_1 \bg_2 = \bg_2 \bg_1 = \bg_0$,
where the group operation is addition modulo $E_6$.

The Voronoi region of $E_6$ has $72$ facets and $54$ vertices. All
vertices have the same norm and hence constitute deep holes of the
lattice.  These holes all lie in $E_6^*$: $27$ are of type $\bg_1$ and
$27$ are of type $\bg_2$.

From the multiplication table, it follows immediately that no
proper subset $\Gamma \subset E_6^*/E_6 = \{\bg_0, \bg_1, \bg_2\}$
except $\{\bg_0\}$ is a group.  Thus, for $L=E_6$, the construction
\eqref{e:ltilde} yields no lattices other than $E_6$ and $E_6^*$.
The NSMs of $E_6$
and $E_6^*$ are $G_{E_6} \triangleq 5/(56\cdot 3^{1/6}) \approx
0.074347$ \cite{conway82voronoi} and $G_{E_6^*} \triangleq
12\,619/(68\,040\cdot 3^{5/6}) \approx 0.074244$ \cite{worley87},
respectively.

The product lattice $E_6 \times E_6$ has nine glue words, formed by 
all Cartesian products $\bg_{ij} \triangleq [\bg_i \; \bg_j]$ for $i,j =
0,1,2$.
The Voronoi region of $E_6 \times E_6$ has $2\cdot 72 = 144$
facets and $54^2 = 2916$ vertices. All these vertices are lattice
vectors in $E^*_6\times E^*_6$ and are of four different types,
namely, the glue words $\bg_{ij}$ for $i,j = 1,2$.

We consider all subsets of the nine glue words $(E_6^*/E_6) \times
(E_6^*/E_6)$ such that $\tilde{L}$ is a lattice, i.e., subsets $\Gamma$
that are groups under addition modulo $E_6 \times E_6$.  From the
group multiplication table it is seen that there are six such
subsets, which are listed in Table~\ref{t:e6e6}. Each of them
generates a lattice $\tilde{L}$ via \eqref{e:ltilde}.

We estimate the NSMs $G$ of the obtained lattices by Monte Carlo
integration using $10^7$ independent samples in the Voronoi region of
each lattice. To find the lattice vector closest to an arbitrary
vector in $\R^{12}$, which is an essential step in generating samples
in the Voronoi regions, we use \cite[Algorithm 5]{ghasemmehdi11}. The
estimated NSMs are presented in the form $\hat{G} \pm 2\hat{\sigma}$,
where $\hat{G}$ is an unbiased estimate of $G$ computed as in
\cite[Eq.~(2)]{conway84} and $\hat{\sigma}$ is an estimate of the
standard deviation of $\hat{G}$ computed as in
\cite[Eq.~(15)]{pook-kolb23}. Because some of the constructed lattices
are \emph{equivalent} to each other by rotation and/or reflection,
there are only four lattices in the table whose geometric properties
such as the NSM differ.

Four of the six groups $\Gamma$ are direct products of groups of $6$-dimensional
vectors, namely, $\{\bg_0\}$ and/or $\{\bg_0,\bg_1,\bg_2\}$,
so the corresponding
lattices are product lattices. The NSMs of $E_6\times
E_6$ and $E_6^*\times E_6^*$ are $G_{E_6}$ and $G_{E_6^*}$, which are
defined above. The NSM of $E_6\times E_6^*$ and $E_6^*\times E_6$ can be calculated from
\cite[Prop.~3]{agrell23} as $(3^{1/6}G_{E_6}+3^{-1/6}G_{E_6^*})/2 =
7711/102\,060 \approx 0.075554$. These exact NSMs are also shown in
Table~\ref{t:e6e6}.

As discussed in Section~\ref{s:fundamentals}, product lattices cannot
be optimal quantizers, so it comes as no surprise that the only
nonproduct lattice in Table~\ref{t:e6e6} is also the best
quantizer. This lattice is obtained by applying \eqref{e:ltilde} to
$L=E_6\times E_6$ with $\Gamma = \{\bg_{00},\bg_{11},\bg_{22}\}$ or
equivalently $\{\bg_{00},\bg_{12},\bg_{21}\}$. What \emph{is}
surprising is that its estimated NSM $\hat{G} \approx 0.070060$ is
\emph{below} the NSM of the Coxeter--Todd lattice $K_{12}$, which was
suggested for quantization by Conway and Sloane in 1984
\cite{conway84} and has remained unsurpassed since
then. The exact NSM of $K_{12}$ was computed in \cite{dutour09} and
is $797\,361\,941/(6\,567\,561\,000 \sqrt{3}) \approx 0.070096$.

To confirm the record, we investigated the new lattice
analytically. A generator matrix of an equivalent lattice is
\eqlab{e:e6e6}{
  \left[\begin{array}{c@{\;\;}c@{\;\;}c@{\;\;}c@{\;\;}c@{\;\;}c@{\;\;}c@{\;\;}c@{\;\;}c@{\;\;}c@{\;\;}c@{\;\;}c}
      2 & 0 & 0 & 0 & 0 & 0 & 0 & 0 & 0 & 0 & 0 & 0 \\ 1 & 1 & 0 & 0 &
      0 & 0 & 0 & 0 & 0 & 0 & 0 & 0 \\ 1 & 0 & 1 & 0 & 0 & 0 & 0 & 0 &
      0 & 0 & 0 & 0 \\ 1 & 0 & 0 & 1 & 0 & 0 & 0 & 0 & 0 & 0 & 0 & 0
      \\ 1 & 0 & 0 & 0 & 1 & 0 & 0 & 0 & 0 & 0 & 0 & 0 \\ \h & \h & \h
      & \h & \h & \tfrac{\sqrt{3}}{2} & 0 & 0 & 0 & 0 & 0 & 0 \\ 0 & 0
      & 0 & 0 & 0 & 0 & 2 & 0 & 0 & 0 & 0 & 0 \\ 0 & 0 & 0 & 0 & 0 & 0
      & 1 & 1 & 0 & 0 & 0 & 0 \\ 0 & 0 & 0 & 0 & 0 & 0 & 1 & 0 & 1 & 0
      & 0 & 0 \\ 0 & 0 & 0 & 0 & 0 & 0 & 1 & 0 & 0 & 1 & 0 & 0 \\ 0 &
      0 & 0 & 0 & 0 & 0 & 1 & 0 & 0 & 0 & 1 & 0 \\ 0 & 0 & 0 & 0 & 0 &
      \tfrac{2}{\sqrt{3}} & -\h & -\h & -\h & -\h & -\h &
      \tfrac{1}{2\sqrt{3}}
\end{array}\right]
,}
which was obtained as described after \eqref{e:ltilde} followed by
negation of some columns for cosmetic reasons.
Using the same method as in \cite{pook-kolb23}, which builds
  upon \cite{Pook-Kolb2022Exact}, the face hierarchy of the Voronoi
region was fully determined. It has in total $11\,432\,765\,485$ faces in dimensions $0$ through $12$, of which $1$ is in dimension $12$ (the Voronoi region
itself), $1602$ are in dimension $11$ (facets), and $65\,665\,350$ are
in dimension $0$ (vertices). The faces lie
  in $702$ equivalence classes under the action of the lattice's symmetry
  group, which has order $10\,749\,542\,400$.  Using the methods
in \cite{pook-kolb23, Pook-Kolb2022Exact}, the exact NSM of the new lattice
is determined to be
\eqlab{e:nsme6e6}{
  G = \frac{200\,359\,601\,790\,277}{2\,859\,883\,842\,816\,000}
  \approx 0.070\,058\,650
,}
confirming the numerical estimate in Table~\ref{t:e6e6}.
Its covariance matrix is proportional
to the identity, which is a necessary but not sufficient condition for
global and local optimality \cite{zamir96,agrell23}.
The complete face catalog is available online \cite[Ancillary files]{face-catalogs12d}.

In comparison with $K_{12}$, which has a symmetry group of order
  $78\,382\,080$ \cite[p.~129]{conway99splag}, the lattice
  \eqref{e:e6e6} is much more symmetric with a symmetry group whose
  order is about $137$ times as large.  Despite this, with
  $11\,971\,901\,593$ faces in $809$ classes (determined with the
  methods in \cite{pook-kolb23}), $K_{12}$ has about $5\,\%$ more
  faces than \eqref{e:e6e6}.

The Voronoi region inherits some properties from (unglued) $E_6\times E_6$, having the same packing radius $1/\sqrt{2}$
and kissing number $144$. It has however three times the packing density, which is $0.02086$. Its covering radius is $2/\sqrt{3}$, which is a factor $\sqrt{2}$ less than the covering radius of $E_6\times E_6$, and its thickness is $7.502$.
The lattice is
equivalent to its dual, but it is not strictly self-dual, because the
dual is a rotated version of the lattice itself.

\subsection[]{Glued $D_6\times D_6$}

A generator matrix for the integer lattice $D_6$ is
\eq{
\left[\begin{array}{cccccc}
2 & 0 & 0 & 0 & 0 & 0 \\
1 & 1 & 0 & 0 & 0 & 0 \\
1 & 0 & 1 & 0 & 0 & 0 \\
1 & 0 & 0 & 1 & 0 & 0 \\
1 & 0 & 0 & 0 & 1 & 0 \\
1 & 0 & 0 & 0 & 0 & 1
\end{array}\right]
.}
Since $\det\bA = (\det\bB)^2 = 4$, the lattice has four glue
vectors $D_6^*/D_6$. These can be taken as
\cite[p.~117]{conway99splag}
\eq{
\begin{array}{c@{\,}c@{}cc@{}cc@{}cc@{}cc@{}cc@{}c@{\,}c}
\bg_0=[ && 0 && 0 && 0 && 0 && 0 && 0 & ], \\
\bg_1=\big[ &&\h &&\h &&\h &&\h &&\h &&\h & \big], \\
\bg_2=[ && 1 && 0 && 0 && 0 && 0 && 0 & ], \\
\bg_3=\big[ &-&\h &&\h &&\h &&\h &&\h &&\h & \big] .
\end{array}
}
This glue group is the point group $C_\mathrm{2v}$, which has order
four. The group identity is $\bg_0$, and the multiplication table
reads $\bg_1^2 = \bg_2^2 = \bg_3^2 = \bg_0$, $\bg_1 \bg_2 = \bg_3$,
$\bg_1 \bg_3 = \bg_2$, and $\bg_2 \bg_3 = \bg_1$, where the (Abelian)
group operation is addition modulo $D_6$.

The Voronoi region of $D_6$ has $60$ facets and $76$ vertices, which
are elements of $D_6^*$.  The vertices consist of $64$ deep holes,
which are of type $\bg_1$ or $\bg_3$, and $12$ shallow holes of type
$\bg_2$.
Setting $L=D_6$ in \eqref{e:ltilde}, the glued lattice
$\tilde{L}$ is one of $D_6$, $D^+_6$, $\Z^6$, or $D^*_6$, depending on
the choice of $\Gamma$. In addition, \eqref{e:ltilde} yields
several nonlattice packings if $\Gamma$ is not a group under addition modulo
$D_6$.

The product lattice $L=D_6 \times D_6$ has $16$ glue words, which are
all Cartesian products $\bg_{ij} \triangleq [\bg_i \; \bg_j]$ for $i,j =
0,1,2,3$.
Its Voronoi region has $2\cdot 60 = 120$
facets and $76^2 = 5776$ vertices. All vertices are lattice vectors in
$D^*_6\times D^*_6$ and are of nine different types, namely, the glue
words $\bg_{ij}$ for $i,j = 1,2,3$.

We consider all subsets $\Gamma$ of the glue words
$(D_6^*/D_6) \times (D_6^*/D_6)$
such that $\tilde{L}$ in \eqref{e:ltilde} is a lattice, i.e., subsets that are
groups under addition modulo $L$. There are $67$ such
subsets. However, several of these subsets are equivalent to each
other, in the sense that a rotation and/or reflection operation in the
symmetry group of $L$ transforms all elements of one
subset into the elements of another subset. The relevant symmetry
operations are (i) interchanging $\bg_{ij}$ with $\bg_{ji}$ throughout
$\Gamma$, i.e., swapping the first set of $6$ coordinates with the last
$6$; (ii) replacing all occurrences of $\bg_{1j}$ with $\bg_{3j}$ and
vice versa; and (iii) replacing all occurrences of $\bg_{i1}$ with
$\bg_{i3}$ and vice versa. If only inequivalent subsets of glue words
are considered, $22$ subsets remain. We furthermore exclude the $10$
subsets that are direct products of $6$-dimensional glue groups, which
generate product lattices. The remaining $12$ glue groups $\Gamma$
are listed in Table~\ref{t:d6d6}. The NSM of each corresponding
lattice $\tilde{L}$ was estimated as in Section~\ref{s:e6e6}.

\begin{table}
\caption{Lattices $\tilde{L}$ generated by gluing $L = D_6\times D_6$ according to \eqref{e:ltilde}. Product lattices and multiple occurrences of equivalent lattices are excluded.}
\label{t:d6d6}
\begin{center}
\renewcommand\arraystretch{1.3}  %
\begin{tabular}{ll}
\hline
Glue words $\Gamma$ & Estimated NSM of $\tilde{L}$ \\
\hline\hline
$\{\bg_{00},\bg_{11}\}$ & $0.071771 \pm 0.000008$ \\
$\{\bg_{00},\bg_{12}\}$ & $0.074092 \pm 0.000010$ \\
$\{\bg_{00},\bg_{22}\}$ & $0.077095 \pm 0.000012$ \\
$\{\bg_{00},\bg_{01},\bg_{12},\bg_{13}\}$ & $0.072099 \pm 0.000008$ \\
$\{\bg_{00},\bg_{01},\bg_{22},\bg_{23}\}$ & $0.075170 \pm 0.000010$ \\
$\{\bg_{00},\bg_{02},\bg_{11},\bg_{13}\}$ & $0.073558 \pm 0.000009$ \\
$\{\bg_{00},\bg_{02},\bg_{21},\bg_{23}\}$ & $0.075909 \pm 0.000010$ \\
$\{\bg_{00},\bg_{11},\bg_{22},\bg_{33}\}$ & $0.070705 \pm 0.000008$ \\
$\{\bg_{00},\bg_{11},\bg_{23},\bg_{32}\}$ & $0.070034 \pm 0.000007$ \\
$\{\bg_{00},\bg_{01},\bg_{10},\bg_{11},\bg_{22},\bg_{23},\bg_{32},\bg_{33}\}$ & $0.071753 \pm 0.000008$ \\
$\{\bg_{00},\bg_{01},\bg_{12},\bg_{13},\bg_{20},\bg_{21},\bg_{32},\bg_{33}\}$ & $0.072887 \pm 0.000008$ \\
$\{\bg_{00},\bg_{02},\bg_{11},\bg_{13},\bg_{20},\bg_{22},\bg_{31},\bg_{33}\}$ & $0.074801 \pm 0.000009$ \\
\hline
\end{tabular}
\end{center}
\end{table}

Numerical studies indicated that one lattice stands out among the dozen new lattices. When $\Gamma =
\{\bg_{00},\bg_{11},\bg_{23},\bg_{32}\}$, we estimate an NSM of $\hat{G} \approx 0.070034$,
which is the smallest value reported to date for $12$-dimensional
lattices. It is slightly smaller than for the best $E_6\times
E_6$-based lattice in Section~\ref{s:e6e6}, and therefore also better
than $K_{12}$.

With this as motivation, we again investigated the new lattice analytically. A generator
matrix of an equivalent lattice is
\eqlab{e:d6d6}{
  \left[\begin{array}{c@{\;\;}c@{\;\;}c@{\;\;}c@{\;\;}c@{\;\;}c@{\;\;}c@{\;\;}c@{\;\;}c@{\;\;}c@{\;\;}c@{\;\;}c}
      2 & 0 & 0 & 0 & 0 & 0 & 0 & 0 & 0 & 0 & 0 & 0 \\
      1 & 1 & 0 & 0 & 0 & 0 & 0 & 0 & 0 & 0 & 0 & 0 \\
      1 & 0 & 1 & 0 & 0 & 0 & 0 & 0 & 0 & 0 & 0 & 0 \\
      1 & 0 & 0 & 1 & 0 & 0 & 0 & 0 & 0 & 0 & 0 & 0 \\
      1 & 0 & 0 & 0 & 1 & 0 & 0 & 0 & 0 & 0 & 0 & 0 \\
      1 & 0 & 0 & 0 & 0 & 1 & 0 & 0 & 0 & 0 & 0 & 0 \\
      \h & \h & \h & \h & \h & \h & 1 & 0 & 0 & 0 & 0 & 0 \\
      0 & 0 & 0 & 0 & 0 & 0 & 1 & 1 & 0 & 0 & 0 & 0 \\
      0 & 0 & 0 & 0 & 0 & 0 & 1 & 0 & 1 & 0 & 0 & 0 \\
      0 & 0 & 0 & 0 & 0 & 0 & 1 & 0 & 0 & 1 & 0 & 0 \\
      0 & 0 & 0 & 0 & 0 & 0 & 1 & 0 & 0 & 0 & 1 & 0 \\
      1 & 0 & 0 & 0 & 0 & 0 & \h & \h & \h & \h & \h & \h
\end{array}\right]
.}
Using the same method as before, it was found that the Voronoi region has $1912$ facets, $21\,273\,456$ vertices, and
$10\,395\,549\,553$ faces overall. The faces fall into $2\,542$ equivalence classes
under its symmetry group, which has order $1\,061\,683\,200$.
The face catalog is available online \cite[Ancillary files]{face-catalogs12d}.
The NSM of the new lattice is
\eq{
G = \frac{6\,492\,178\,537\,549}{92\,704\,053\,657\,600} \approx 0.070\,031\,226
}
and the covariance matrix is proportional to the identity.

The packing radius and kissing number of the lattice \eqref{e:d6d6} are $1/\sqrt{2}$ and $120$, respectively, which are the same as for $D_6\times D_6$. The packing density is $0.02086$, four times larger than for $D_6\times D_6$ and the same as for the lattice \eqref{e:e6e6}. The covering radius is $\sqrt{3/2}$, which is a factor of $\sqrt{2}$ smaller than for $D_6\times D_6$, and the thickness is $15.21$. Like \eqref{e:e6e6}, the lattice generated by \eqref{e:d6d6} is
equivalent to its own dual.

\section{Conclusions}

Contrary to the common belief,
there exist lattices
with lower second moments than $K_{12}$.
One such lattice
is a union of three translated copies of $E_6\times E_6$ and another
is a union of four translated copies of $D_6\times D_6$. The latter
sets a new record for $12$-dimensional lattice quantizers.

As a byproduct, an improved $13$-dimensional lattice quantizer is obtained.
The previously best published lattice for $n=13$ is a product of $K_{12}$ and
a scaled integer lattice $a\Z$, with an NSM of $0.071035$ \cite[Tab.~I]{lyu22}.
Replacing $K_{12}$ with the new best $12$-dimensional lattice in a similar
product construction yields a slightly improved NSM of $0.070974$
(which is however inferior to a yet unpublished laminated lattice \cite{Pook-Kolb2022Exact}).

Applying gluing theory to the design of lattice quantizers clearly holds
great promise, and can probably lead to better quantizers in other dimensions as well.
Even in $12$ dimensions, the fundamental question remains open: Can even better lattice quantizers
be found, by gluing theory or other methods?

\balance


\begin{thebibliography}{10}
\providecommand{\url}[1]{#1}
\csname url@samestyle\endcsname
\providecommand{\newblock}{\relax}
\providecommand{\bibinfo}[2]{#2}
\providecommand{\BIBentrySTDinterwordspacing}{\spaceskip=0pt\relax}
\providecommand{\BIBentryALTinterwordstretchfactor}{4}
\providecommand{\BIBentryALTinterwordspacing}{\spaceskip=\fontdimen2\font plus
\BIBentryALTinterwordstretchfactor\fontdimen3\font minus
  \fontdimen4\font\relax}
\providecommand{\BIBforeignlanguage}[2]{{%
\expandafter\ifx\csname l@#1\endcsname\relax
\typeout{** WARNING: IEEEtran.bst: No hyphenation pattern has been}%
\typeout{** loaded for the language `#1'. Using the pattern for}%
\typeout{** the default language instead.}%
\else
\language=\csname l@#1\endcsname
\fi
#2}}
\providecommand{\BIBdecl}{\relax}
\BIBdecl

\bibitem{conway99splag}
\BIBentryALTinterwordspacing
J.~H. Conway and N.~J.~A. Sloane, \emph{Sphere Packings, Lattices and Groups},
  3rd~ed.\hskip 1em plus 0.5em minus 0.4em\relax New York, NY: Springer, 1999.
  [Online]. Available: \url{https://doi.org/10.1007/978-1-4757-6568-7}
\BIBentrySTDinterwordspacing

\bibitem{gray98}
\BIBentryALTinterwordspacing
R.~M. Gray and D.~L. Neuhoff, ``Quantization,'' \emph{{IEEE} Trans. Inf.
  Theory}, vol.~44, no.~6, pp. 2325--2383, Oct. 1998. [Online]. Available:
  \url{https://doi.org/10.1109/18.720541}
\BIBentrySTDinterwordspacing

\bibitem{zamir14book}
\BIBentryALTinterwordspacing
R.~Zamir, \emph{Lattice Coding for Signals and Networks}.\hskip 1em plus 0.5em
  minus 0.4em\relax Cambridge, UK: Cambridge University Press, 2014. [Online].
  Available: \url{https://doi.org/10.1017/CBO9781139045520}
\BIBentrySTDinterwordspacing

\bibitem{forney89a}
\BIBentryALTinterwordspacing
G.~D. Forney, Jr. and L.-F. Wei, ``Multidimensional constellations---part {I}:
  Introduction, figures of merit, and generalized cross constellations,''
  \emph{{IEEE} J. Sel. Areas Commun.}, vol.~7, no.~6, pp. 877--892, Aug. 1989.
  [Online]. Available: \url{https://doi.org/10.1109/49.29611}
\BIBentrySTDinterwordspacing

\bibitem{hamprecht03}
\BIBentryALTinterwordspacing
F.~A. Hamprecht and E.~Agrell, ``Exploring a space of materials: Spatial
  sampling design and subset selection,'' in \emph{Experimental Design for
  Combinatorial and High Throughput Materials Development}, J.~N. Cawse,
  Ed.\hskip 1em plus 0.5em minus 0.4em\relax New York, NY: Wiley, 2003, ch.~13.
  [Online]. Available:
  \url{https://citeseerx.ist.psu.edu/viewdoc/download?doi=10.1.1.20.6178}
\BIBentrySTDinterwordspacing

\bibitem{allen21}
\BIBentryALTinterwordspacing
B.~Allen and E.~Agrell, ``The optimal lattice quantizer in nine dimensions,''
  \emph{Annalen der Physik}, vol. 533, no.~12, p. 2100259, Dec. 2021. [Online].
  Available: \url{https://doi.org/10.1002/andp.202100259}
\BIBentrySTDinterwordspacing

\bibitem{fejestoth59}
\BIBentryALTinterwordspacing
L.~{Fejes T\'oth}, ``Sur la repr\'esentation d'une population infinie par un
  nombre fini d'\'el\'ements,'' \emph{Acta Mathematica Hungarica}, vol.~10, no.
  3--4, pp. 299--304, Sept. 1959, in French. [Online]. Available:
  \url{https://doi.org/10.1007/bf02024494}
\BIBentrySTDinterwordspacing

\bibitem{barnes83}
\BIBentryALTinterwordspacing
E.~S. Barnes and N.~J.~A. Sloane, ``The optimal lattice quantizer in three
  dimensions,'' \emph{SIAM J. Alg. Disc. Meth.}, vol.~4, no.~1, pp. 30--41,
  Mar. 1983. [Online]. Available: \url{https://doi.org/10.1137/0604005}
\BIBentrySTDinterwordspacing

\bibitem{gersho79}
\BIBentryALTinterwordspacing
A.~Gersho, ``Asymptotically optimal block quantization,'' \emph{{IEEE} Trans.
  Inf. Theory}, vol. IT-25, no.~4, pp. 373--380, July 1979. [Online].
  Available: \url{https://doi.org/10.1109/TIT.1979.1056067}
\BIBentrySTDinterwordspacing

\bibitem{conway82voronoi}
\BIBentryALTinterwordspacing
J.~H. Conway and N.~J.~A. Sloane, ``Voronoi regions of lattices, second moments
  of polytopes, and quantization,'' \emph{{IEEE} Trans. Inf. Theory}, vol.
  IT-28, no.~2, pp. 211--226, Mar. 1982. [Online]. Available:
  \url{https://doi.org/10.1109/TIT.1982.1056483}
\BIBentrySTDinterwordspacing

\bibitem{agrell98}
\BIBentryALTinterwordspacing
E.~Agrell and T.~Eriksson, ``Optimization of lattices for quantization,''
  \emph{{IEEE} Trans. Inf. Theory}, vol.~44, no.~5, pp. 1814--1828, Sept. 1998.
  [Online]. Available: \url{https://doi.org/10.1109/18.705561}
\BIBentrySTDinterwordspacing

\bibitem{conway84}
\BIBentryALTinterwordspacing
J.~H. Conway and N.~J.~A. Sloane, ``On the {Voronoi} regions of certain
  lattices,'' \emph{SIAM J. Alg. Disc. Meth.}, vol.~5, no.~3, pp. 294--305,
  Sept. 1984. [Online]. Available: \url{https://doi.org/10.1137/0605031}
\BIBentrySTDinterwordspacing

\bibitem{dutour09}
\BIBentryALTinterwordspacing
M.~{Dutour Sikiri\'c}, A.~Sch\"urmann, and F.~Vallentin, ``Complexity and
  algorithms for computing {Voronoi} cells of lattices,'' \emph{Mathematics of
  Computation}, vol.~78, no. 267, pp. 1713--1731, July 2009. [Online].
  Available: \url{https://doi.org/10.1090/S0025-5718-09-02224-8}
\BIBentrySTDinterwordspacing

\bibitem{torquato10}
\BIBentryALTinterwordspacing
S.~Torquato, ``Reformulation of the covering and quantizer problems as ground
  states of interacting particles,'' \emph{Phys. Rev. E}, vol.~82, no.~5, p.
  056109, Nov. 2010. [Online]. Available:
  \url{https://doi.org/10.1103/PhysRevE.82.056109}
\BIBentrySTDinterwordspacing

\bibitem{lyu22}
\BIBentryALTinterwordspacing
S.~Lyu, Z.~Wang, C.~Ling, and H.~Chen, ``Better lattice quantizers constructed
  from complex integers,'' \emph{{IEEE} Trans. Commun.}, vol.~70, no.~12, pp.
  7932--7940, Dec. 2022. [Online]. Available:
  \url{https://doi.org/10.1109/TCOMM.2022.3215685}
\BIBentrySTDinterwordspacing

\bibitem{agrell23}
\BIBentryALTinterwordspacing
E.~Agrell and B.~Allen, ``On the best lattice quantizers,'' \emph{{IEEE} Trans.
  Inf. Theory}, vol.~69, no.~12, pp. 7650--7658, Dec. 2023. [Online].
  Available: \url{https://doi.org/10.1109/TIT.2023.3291313}
\BIBentrySTDinterwordspacing

\bibitem{coxeter53}
\BIBentryALTinterwordspacing
H.~S.~M. Coxeter and J.~A. Todd, ``An extreme duodenary form,'' \emph{Canadian
  Journal of Mathematics}, vol.~5, pp. 384--392, 1953. [Online]. Available:
  \url{https://doi.org/10.4153/CJM-1953-043-4}
\BIBentrySTDinterwordspacing

\bibitem{conway83mpcps}
\BIBentryALTinterwordspacing
J.~H. Conway and N.~J.~A. Sloane, ``The {Coxeter--Todd} lattice, the {Mitchell}
  group, and related sphere packings,'' \emph{Mathematical Proceedings of the
  Cambridge Philosophical Society}, vol.~93, no.~3, pp. 421--440, May 1983.
  [Online]. Available: \url{https://doi.org/10.1017/S0305004100060746}
\BIBentrySTDinterwordspacing

\bibitem{conway79}
\BIBentryALTinterwordspacing
J.~H. Conway, V.~Pless, and N.~J.~A. Sloane, ``Self-dual codes over {GF}(3) and
  {GF}(4) of length not exceeding 16,'' \emph{{IEEE} Trans. Inf. Theory}, vol.
  IT-25, no.~3, pp. 312--322, May 1979. [Online]. Available:
  \url{https://doi.org/10.1109/TIT.1979.1056047}
\BIBentrySTDinterwordspacing

\bibitem{conway82jnt}
\BIBentryALTinterwordspacing
J.~H. Conway and N.~J.~A. Sloane, ``On the enumeration of lattices of
  determinant one,'' \emph{Journal of Number Theory}, vol.~15, no.~1, pp.
  83--94, Aug. 1982. [Online]. Available:
  \url{https://doi.org/10.1016/0022-314X(82)90084-1}
\BIBentrySTDinterwordspacing

\bibitem{conway82ejc}
\BIBentryALTinterwordspacing
------, ``The unimodular lattices of dimension up to 23 and the
  {Minkowski--Siegel} mass constants,'' \emph{European Journal of
  Combinatorics}, vol.~3, no.~3, pp. 219--231, Sept. 1982. [Online]. Available:
  \url{https://doi.org/10.1016/S0195-6698(82)80034-6}
\BIBentrySTDinterwordspacing

\bibitem{gannon91}
\BIBentryALTinterwordspacing
T.~Gannon and C.~S. Lam, ``Gluing and shifting lattice constructions and
  rational equivalence,'' \emph{Reviews in Mathematical Physics}, vol.~3,
  no.~3, pp. 331--369, 1991. [Online]. Available:
  \url{https://doi.org/10.1142/S0129055X91000126}
\BIBentrySTDinterwordspacing

\bibitem{coxeter51}
\BIBentryALTinterwordspacing
H.~S.~M. Coxeter, ``Extreme forms,'' \emph{Canadian Journal of Mathematics},
  vol.~3, pp. 391--441, 1951. [Online]. Available:
  \url{https://doi.org/10.4153/CJM-1951-045-8}
\BIBentrySTDinterwordspacing

\bibitem{worley87}
\BIBentryALTinterwordspacing
R.~T. Worley, ``The {Voronoi} region of {$E^*_6$},'' \emph{J. Austral. Math.
  Soc. (Series A)}, vol.~43, no.~2, pp. 268--278, Oct. 1987. [Online].
  Available: \url{https://doi.org/10.1017/S1446788700029402}
\BIBentrySTDinterwordspacing

\bibitem{ghasemmehdi11}
\BIBentryALTinterwordspacing
A.~Ghasemmehdi and E.~Agrell, ``Faster recursions in sphere decoding,''
  \emph{{IEEE} Trans. Inf. Theory}, vol.~57, no.~6, pp. 3530--3536, June 2011.
  [Online]. Available: \url{https://doi.org/10.1109/TIT.2011.2143830}
\BIBentrySTDinterwordspacing

\bibitem{pook-kolb23}
\BIBentryALTinterwordspacing
D.~Pook-Kolb, E.~Agrell, and B.~Allen, ``The {Voronoi} region of the
  {Barnes--Wall} lattice {$\Lambda_{16}$},'' \emph{{IEEE} J. Sel. Areas Inf.
  Theory}, vol.~4, pp. 16--23, 2023. [Online]. Available:
  \url{https://doi.org/10.1109/JSAIT.2023.3276897}
\BIBentrySTDinterwordspacing

\bibitem{Pook-Kolb2022Exact}
\BIBentryALTinterwordspacing
D.~Pook-Kolb, B.~Allen, and E.~Agrell, ``Exact calculation of quantizer
  constants for arbitrary lattices,'' arXiv preprint, 2022--2024. [Online].
  Available: \url{https://arxiv.org/abs/2211.01987}
\BIBentrySTDinterwordspacing

\bibitem{zamir96}
\BIBentryALTinterwordspacing
R.~Zamir and M.~Feder, ``On lattice quantization noise,'' \emph{{IEEE} Trans.
  Inf. Theory}, vol.~42, no.~4, pp. 1152--1159, July 1996. [Online]. Available:
  \url{https://doi.org/10.1109/18.508838}
\BIBentrySTDinterwordspacing

\bibitem{face-catalogs12d}
\BIBentryALTinterwordspacing
E.~Agrell, D.~Pook-Kolb, and B.~Allen, ``Glued lattices are better quantizers
  than {$K_{12}$},'' arXiv preprint and data, Dec.~2023. [Online]. Available:
  \url{http://arxiv.org/abs/2312.00481}
\BIBentrySTDinterwordspacing

\end{thebibliography}
\end{document}